\documentclass[Floatfix, a4paper, amsfonts, amssymb, amsmath, reprint, showkeys, nofootinbib, twocolumn, twoside, aps, 10pt ]{revtex4}

\usepackage{natbib}
\usepackage{mathtools}
\usepackage{scalerel}

\usepackage{tikz}
\usetikzlibrary{svg.path}
\definecolor{kjm}{HTML}{A6CE39}
\tikzset{
    kz/.pic={
        \fill[kjm] svg{M256,128c0,70.7-57.3,128-128,128C57.3,256,0,198.7,0,128C0,57.3,57.3,0,128,0C198.7,0,256,57.3,256,128z};
        \fill[white] svg{M86.3,186.2H70.9V79.1h15.4v48.4V186.2z}
        svg{M108.9,79.1h41.6c39.6,0,57,28.3,57,53.6c0,27.5-21.5,53.6-56.8,53.6h-41.8V79.1z M124.3,172.4h24.5c34.9,0,42.9-26.5,42.9-39.7c0-21.5-13.7-39.7-43.7-39.7h-23.7V172.4z}
        svg{M88.7,56.8c0,5.5-4.5,10.1-10.1,10.1c-5.6,0-10.1-4.6-10.1-10.1c0-5.6,4.5-10.1,10.1-10.1C84.2,46.7,88.7,51.3,88.7,56.8z};
    }
}
\newcommand\axc[1]{\href{https://orcid.org/#1}{\mbox{\scalerel*{
                \begin{tikzpicture}[yscale=-1,transform shape]
                \pic{kz};
                \end{tikzpicture}
            }{|}}}}

\usepackage[english]{babel}
\usepackage[utf8]{inputenc}
\usepackage[colorinlistoftodos, color=green!40, prependcaption]{todonotes}
\usepackage{amsthm}
\usepackage{mathtools}
\usepackage{xcolor}
\usepackage{graphicx}
\usepackage[left=23mm,right=13mm,top=35mm,columnsep=15pt]{geometry} 
\usepackage{adjustbox}
\usepackage{lipsum}
\usepackage{csquotes}
\usepackage{subfigure}
\usepackage{placeins}
\usepackage{float}
\usepackage{mathalpha}
\usepackage{slashed}
\usepackage[pdftex, pdftitle={Article}, pdfauthor={Author}]{hyperref} % For hyperlinks in the PDF
\begin{document}

\title{Infrared Lines from Sterile-Neutrino Transition Magnetic Moments at JWST}

\author{Hriditi Howlader$^{\axc{0009-0007-9167-3487}}$\,}
\email{p22ph005@nitm.ac.in}
 \affiliation{Department of Physics, National Institute of Technology Meghalaya, Shillong, Meghalaya, India}

\author{Alekha C. Nayak$^{\axc{0000-0001-6087-2490}}$\,}
    \email{alekhanayak@nitm.ac.in}
    \affiliation{Department of Physics, National Institute of Technology Meghalaya, Shillong, Meghalaya, India}

\author{Tripurari Srivastava$^{\axc{0000-0001-6856-9517}}$\,}
    \email{tripurarisri022@gmail.com}
    \affiliation{Department of Physics and Astrophysics, University of Delhi, Delhi-110007, Delhi, India}
   %  \affiliation{Institute of Particle Physics and Key Laboratory of Quark and Lepton Physics (MOE),
%Central China Normal University, Wuhan, Hubei 430079, China}

\begin{abstract}

We investigate infrared line signatures from radiatively decaying
sterile-neutrino dark matter using publicly available JWST/NIRSpec
IFU blank-sky observations. The main signal considered is the
sterile-to-sterile transition $N_1\to N_2\gamma$, induced by the
transition magnetic dipole coefficient $d_{NN\gamma}$, with
$m_1>m_2$. In contrast to ordinary two-photon decays of axion-like
or Majoron-like particles, the observed photon energy is not fixed by
the full dark matter mass, but by the small mass splitting
$\Delta m=m_1-m_2$. Thus, a keV-scale sterile-neutrino dark matter
state can generate an eV-scale infrared photon line in the JWST band.
We construct a $\chi^2$-based line-search analysis using the
NIRSpec IFU $\rm F170LP$-$\rm G235M$ blank-sky data toward
$\rm GN\text{-}z11$, modelling the smooth continuum with a cubic
spline and including the Milky Way halo decay flux. In the absence of
a significant excess, we derive projected limits on $d_{NN\gamma}$
and on the decay width $\Gamma_{N_1\to N_2\gamma}$ for
$0.1~{\rm eV}\lesssim\Delta m\lesssim1~{\rm eV}$. For a sterile
component saturating the dark matter abundance, the strongest
sensitivity reaches $d_{NN\gamma}\lesssim7\times10^{-14}~{\rm GeV}^{-1}$
and $\Gamma_{N_1\to N_2\gamma}\lesssim10^{-25}~{\rm s}^{-1}$.
We also include a minimal anomalous-Majoron benchmark,
$\omega\to\gamma\gamma$, obtaining JWST sensitivity to
$\lambda_{\omega\gamma\gamma}\sim10^{-11}$-$10^{-9}~{\rm GeV}^{-1}$
for $m_\omega\sim0.6$-$1~{\rm eV}$.

\end{abstract}

\maketitle

%=========================================================
\section{Introduction}
\label{sec:introduction}
%=========================================================

The microscopic nature of dark matter remains one of the central open questions in particle physics and cosmology. Its existence is firmly
supported by a wide range of gravitational observations, including cosmic microwave background measurements, large-scale structure, galaxy rotation curves and gravitational lensing, yet its particle identity remains unknown~\cite{Planck:2018vyg,janish2025hunting}.
Although many dark matter models assume absolute stability protected by an exact symmetry, cosmological viability only requires the dark matter lifetime to exceed the age of the Universe. Very slow dark matter decay into Standard Model particles therefore provides a well-motivated and highly predictive avenue for indirect detection.

Among the cleanest indirect-detection signatures are narrow photon lines from two-body decays ~\cite{janish2025hunting, roy2025sensitivity,pinetti2025first}. Unlike continuum emission, a monochromatic
line has a sharply defined spectral position fixed by the underlying particle masses, and can therefore be searched for against smooth
astrophysical backgrounds. Infrared line searches are particularly interesting for dark sectors with eV-scale photon energies, a region that is inaccessible to conventional X-ray and gamma-ray searches but directly relevant for several light or nearly degenerate dark matter
scenarios.

The James Webb Space Telescope (JWST) has opened a new observational window for such searches. Its infrared spectroscopic instruments, including NIRSpec, cover wavelengths from the near to mid infrared range and provide excellent sensitivity to narrow spectral features over photon energies of order an electronvolt~\cite{Gardner:2006ky,
jakobsen2022near,janish2025hunting}. JWST blank-sky observations have therefore become a powerful dataset for testing decaying dark matter
models through line emission from the Milky Way halo \cite{pinetti2025first}. Existing studies have already demonstrated the utility of JWST spectroscopy for axion-like particle and related pseudoscalar dark matter searches. Here we use the same observational opportunity to probe a different neutrino-sector origin of an infrared line.

The main focus of this work is a quasi-degenerate sterile-neutrino dark sector. Sterile neutrinos are among the simplest extensions of the Standard Model and are broadly motivated by neutrino mass
generation, early-Universe cosmology and dark matter phenomenology \cite{Abazajian:2017tcc}. We consider the case in which a heavier keV-scale sterile state decays into a lighter sterile state and a single photon through a sterile-to-sterile transition magnetic moment~\cite{Beltran:2024twr,Howlader:2025gvz}. In this scenario, the observed photon energy is controlled not by the full sterile-neutrino mass, but by the small mass splitting between the two sterile states. Consequently, a keV-scale dark matter particle can give rise to an eV-scale infrared line when the sterile states are separated by an eV-scale mass splitting. This kinematic feature makes the signal qualitatively distinct from ordinary two-photon decays of axion-like particles or Majoron-like
pseudoscalars.

Sterile-neutrino transition magnetic moments have been studied within low-energy and electroweak effective field theory frameworks, and their ultraviolet completions can involve charged scalars and vector-like charged leptons~\cite{Beltran:2024twr}. Previous
phenomenological studies have mostly focused on heavier sterile neutrinos and long-lived particle signatures at collider and beam-dump experiments. In contrast, the present work investigates the
infrared astrophysical regime, where JWST can test transition magnetic moments through narrow line emission from radiatively decaying sterile-neutrino dark matter.

Using publicly available JWST/NIRSpec Integral Field Unit (IFU) blank-sky observations, we perform a dedicated search for narrow infrared features associated with sterile-neutrino radiative decay in the Milky Way halo. We construct a statistical analysis based on the observed blank-sky spectra and derive upper limits on the sterile-neutrino decay width and on the sterile-to-sterile transition
magnetic dipole coefficient over the mass-splitting range $0.1~{\rm eV}\lesssim\Delta m\lesssim1~{\rm eV}$. For a sterile neutrino component saturating the dark matter abundance, the strongest
limits reach decay widths of order $10^{-25}~{\rm s}^{-1}$ and transition dipole coefficients of order
$10^{-13}$--$10^{-14}~{\rm GeV}^{-1}$ in the most sensitive part of the JWST/NIRSpec wavelength range. These results show that infrared
spectroscopy can provide a competitive and complementary probe of neutrino-sector electromagnetic interactions.

For completeness, we also include a short benchmark discussion of an anomalous Majoron. The Majoron is a pseudo-Nambu-Goldstone boson associated with the spontaneous breaking of global lepton number and
arises naturally in neutrino mass models~\cite{Chikashige:1980qk,
Gelmini:1980re,Georgi:1981pg,Schechter:1981cv}. At low energies, an anomalous Majoron can decay into two photons, producing an infrared line whose phenomenology closely parallels the standard axion-like particle interpretation~\cite{Lu:2025kbe,roy2025sensitivity}. Since this case has substantial overlap with existing pseudoscalar line searches, we treat it as a comparison
benchmark rather than as the central new ingredient of the paper. The effective transition magnetic moment is treated as the primary parameter
constrained by JWST data, while possible ultraviolet completions are discussed only to illustrate how such an operator may arise in
sterile-neutrino extensions of the Standard Model. In particular, quasi-degenerate sterile states may be motivated by approximate lepton-number symmetry or pseudo-Dirac constructions, and the transition dipole may be generated by electrically charged mediators. The detailed model-building assumptions are not required for the main JWST limits.

The paper is organized as follows. In Sec.~\ref{sec:theory_flux}, we
present the theoretical framework and photon-flux formalism used in
this work. This section introduces the sterile-neutrino transition
magnetic moment, the Majoron two-photon benchmark, and the common
infrared line-flux formalism for decaying dark matter in the Milky
Way halo. In Sec.~\ref{sec:results}, we describe the JWST/NIRSpec
blank-sky analysis, the statistical procedure used to obtain limits,
and the resulting constraints on the sterile-to-sterile transition
magnetic coupling strength \(d_{NN\gamma}\) and the decay width
\(\Gamma_{N_1\to N_2\gamma}\). We also include a minimal Majoron
benchmark result, expressed in terms of the Majoron-photon coupling and the decay width. In
Sec.~\ref{sec:UV_interpretation}, we discuss possible ultraviolet
interpretations of the sterile transition dipole, together with a
brief comment on the Majoron benchmark. Finally, we summarize our
conclusions and outlook in Sec.~\ref{sec:Conclusion}.

                                        %=========================================================
\section{Theoretical framework and photon flux}
\label{sec:theory_flux}
%=========================================================

In this section we formulate the two infrared line signals considered in this work. The primary focus is the radiative decay of a quasi-degenerate sterile-neutrino pair through a sterile-to-sterile transition magnetic moment. We also include, as a benchmark, the two-photon decay of an anomalous Majoron. Both signals produce narrow infrared spectral lines and can therefore be constrained using the same JWST line-search strategy, although their particle-physics interpretations are different.

%=========================================================
\subsection{Sterile-neutrino transition magnetic moment}
\label{subsec:sterile_tmm}
%=========================================================

We consider two sterile-neutrino mass eigenstates, $N_1$ and $N_2$,
ordered as
\begin{equation}
m_1>m_2,
\qquad
\Delta m\equiv m_1-m_2 .
\end{equation}
The sterile-neutrino masses are assumed to lie well below the electroweak scale. The relevant electromagnetic interaction can therefore be described within the low-energy $N_R$LEFT framework. In the mass basis, the sterile-to-sterile transition dipole operator is \cite{Beltran:2024twr},
\begin{equation}
\mathcal{L}_{N_R{\rm LEFT}}
\supset
d_{NN\gamma}
\left(
\overline{N_2^{\,c}}\sigma^{\mu\nu}P_RN_1
\right)F_{\mu\nu}
+
{\rm H.c.},
\label{eq:sterile_dipole_operator}
\end{equation}
where $d_{NN\gamma}$ is the transition magnetic dipole coefficient and $F_{\mu\nu}$ is the electromagnetic field-strength tensor \cite{Beltran:2024twr}. For sterile neutrinos, diagonal magnetic moments vanish, while off-diagonal transition moments are allowed.

The operator in Eq.~\eqref{eq:sterile_dipole_operator} induces the radiative decay
\begin{equation}
N_1\rightarrow N_2+\gamma .
\end{equation}
The corresponding decay width is  \cite{Beltran:2024twr},\cite{Howlader:2025gvz},
\begin{equation}
\Gamma_{N_1\rightarrow N_2\gamma}
=
\frac{2|d_{NN\gamma}|^2}{\pi}
m_1^3
\left(
1-\frac{m_2^2}{m_1^2}
\right)^3 .
\label{eq:sterile_decay_width_exact}
\end{equation}
Defining
\begin{equation}
\delta\equiv 1-\frac{m_2}{m_1}
=
\frac{\Delta m}{m_1},
\end{equation}
this can also be written as \cite{Beltran:2024twr}
\begin{equation}
\Gamma_{N_1\rightarrow N_2\gamma}
=
\frac{2|d_{NN\gamma}|^2}{\pi}
m_1^3(2-\delta)^3\delta^3 .
\label{eq:sterile_decay_width_delta}
\end{equation}
In the quasi-degenerate limit, $\Delta m\ll m_1$, one obtains \cite{Howlader:2025gvz}
\begin{equation}
\Gamma_{N_1\rightarrow N_2\gamma}
\simeq
\frac{16|d_{NN\gamma}|^2}{\pi}
(\Delta m)^3 .
\label{eq:sterile_decay_width_degenerate}
\end{equation}

The energy of the emitted photon is
\begin{equation}
E_\gamma
=
\frac{m_1^2-m_2^2}{2m_1}
=
\Delta m
\left(
1-\frac{\Delta m}{2m_1}
\right)
\simeq
\Delta m .
\label{eq:sterile_photon_energy}
\end{equation}
Hence, a keV-scale sterile-neutrino dark matter state can produce an eV-scale infrared photon if the splitting between the two sterile states is of order $0.1$--$1~{\rm eV}$. This is the key distinction from ordinary two-photon decays of axion-like or Majoron-like dark matter, where the photon energy is set by the dark matter mass itself.

%=========================================================
\subsection{Majoron two-photon benchmark}
\label{subsec:majoron_benchmark_theory}
%=========================================================

For comparison, we also consider an anomalous Majoron-like pseudoscalar, denoted by $\omega$, coupled to photons through the effective interaction \cite{arias2021neutrino}
\begin{equation}
\mathcal{L}_{\omega}^{\rm eff}
\supset
\frac{1}{4}\lambda_{\omega\gamma\gamma}
\omega F_{\mu\nu}\widetilde F^{\mu\nu},
\label{eq:majoron_photon_operator}
\end{equation}
where
\begin{equation}
\widetilde F^{\mu\nu}
=
\frac{1}{2}\epsilon^{\mu\nu\rho\sigma}F_{\rho\sigma}.
\end{equation}
In this convention, at low energy, the two-photon decay width of decaying Majorons is \cite{arias2021neutrino}
\begin{equation}
\Gamma_{\omega\rightarrow\gamma\gamma}
=
\frac{
\lambda_{\omega\gamma\gamma}^{2}m_\omega^3
}{32\pi}.
\label{eq:majoron_decay_width}
\end{equation}
The emitted photons have energy
\(
E_\gamma=\frac{m_\omega}{2}.
\label{eq:majoron_photon_energy}
\)
It has been observed that the tension on the Hubble constant is reduced for the mass range of Majorons $m_{\omega}\in [0.1, 1]~\rm eV$ and coupling strength $\lambda_{\omega\nu\nu}\in [5\times 10^{-14}, 10^{-12}]$ \cite{arias2021neutrino}, \cite{escudero2020cmb}. Whereas, for the same Majoron mass range $m_{\omega}\in [0.1, 1]~\rm eV$ and effective coupling strength to photons are set by CAST, whose upper bound limit is $\lambda_{\omega\gamma\gamma}\lesssim 10^{-10}~\rm GeV^{-1}$ \cite{anastassopoulos2017new}.

This signal has the same line phenomenology as the standard axion-like-particle two-photon decay. We therefore use the Majoron case only as a benchmark comparison, while the main new analysis concerns the sterile-neutrino transition magnetic moment.

%=========================================================
\subsection{Photon flux from decaying dark matter}
\label{subsec:general_flux}
%=========================================================

The photon line intensity from decaying dark matter is proportional to the dark matter column density along the line of sight. For an observed direction $(\ell,b)$, we define the effective dark matter density $D$~ factor as \cite{roy2025sensitivity}
\begin{equation}
D_{\rm eff}(\lambda,\ell,b)
=
\int_{\rm l.o.s.} ds\,
\rho_{\rm DM}\!\left[r(s,\ell,b)\right]\,
e^{-\tau(\lambda,s)} ,
\label{eq:Deff}
\end{equation}
where $\rho_{\rm DM}$ is the dark matter density and
$e^{-\tau(\lambda,s)}$ accounts for wavelength-dependent dust extinction between the decay point and the observer. If all the emission originated at infinity, then this exponential term can be written as, $e^{-\tau(\lambda,s)}\longrightarrow 10^{-0.44A_{\lambda}}$ \cite{roy2025sensitivity}. This $A_{\lambda}$ represents the galactic dust extinction at wavelength $\lambda$, and it follows a wavelength-dependent power law \cite{larson2005reddening}. The line-of-sight distance from the Galactic centre is \cite{saha2025shedding}
\begin{equation}
r(s,\ell,b)
=
\sqrt{
s^2+r_\odot^2
-
2sr_\odot\cos\ell\cos b
},
\label{eq:los_distance}
\end{equation}
with $r_\odot=8.1~{\rm kpc}$. In the numerical analysis, we use the Navarro–Frenk–White density profile \cite{Navarro:1995iw}
\begin{equation}
\rho_{\rm NFW}(r)
=
\frac{\rho_s}{
(r/r_s)(1+r/r_s)^2
},
\label{eq:NFW_profile}
\end{equation}
with $\rho_s=0.18~{\rm GeV\,cm^{-3}}$ and
$r_s=24.1~{\rm kpc}$ \cite{Navarro:1996gj,Cirelli:2010xx,
Cirelli:2020bpc,maity2022search,sun2025dynamical}.

For an extended field of view, the averaged effective dark matter density $D$~ factor is \cite{Cirelli:2010xx}
\begin{equation}
\overline D_{\rm eff}(\lambda)
=
\frac{1}{\Delta\Omega}
\int_{\Delta\Omega} d\Omega\,
D_{\rm eff}(\lambda,\ell,b),
\label{eq:averaged_Dfactor}
\end{equation}
where $\Delta\Omega$ is the solid angle of the observed region after applying the relevant masks and instrumental cuts. This definition
should be understood as the actual JWST field of view used in the analysis, rather than a full-sky average.

%he normalized line profile is denoted by $G_\lambda(\lambda)$ and it satisfies
%
%\begin{equation}
%\int d\lambda\,G_\lambda(\lambda)=1.
%\end{equation}
%
We model the observed profile as a Gaussian whose width includes both instrumental resolution and Doppler broadening \cite{saha2025shedding},%\textcolor{red}{the observed Gaussian profile will be multiplied by $2$ as two photons are decaying }
\begin{equation}
G_\lambda(\lambda)
=  \frac{1}{\sqrt{2\pi}\sigma_\lambda}
\exp\left[
-\frac{(\lambda-\lambda_\gamma)^2}
{2\sigma_\lambda^2}
\right],
\label{eq:line_profile}
\end{equation}
with
\begin{equation}
\sigma_\lambda^2
=
\sigma_{\lambda,{\rm inst}}^2
+
\lambda_\gamma^2\frac{\sigma_v^2}{c^2}.
\label{eq:line_width}
\end{equation}
Here $\sigma_v\simeq200~{\rm km\,s^{-1}}$ is taken as a benchmark Milky Way velocity dispersion and $\sigma_{\lambda,\rm inst} = \frac{\Delta \lambda}{2\sqrt{2\rm ln(2)}}$, where $\Delta \lambda$ is connected to the spectral resolution of the instrument \cite{Peter:2013aha,Evans:2018bqy}. The instrumental contribution is determined by the JWST/NIRSpec configuration used in the analysis.

For sterile-neutrino decay, the differential energy intensity is
\begin{equation}
I_N(d_{NN\gamma}) = \frac{dI_N}{d\lambda\,d\Omega}
=
\frac{
f_N\,
\Gamma_{N_1\rightarrow N_2\gamma}
E_{\gamma,N}
}{
4\pi m_1
}
G_\lambda^{N}(\lambda)
D_{\rm eff}(\lambda,\ell,b),
\label{eq:sterile_flux}
\end{equation}
where $f_N$ is the fraction of the dark matter abundance carried by the decaying sterile state and
$E_{\gamma,N}\simeq\Delta m$. $G_\lambda^{N}(\lambda)$ represents the instrumental resolution and Doppler broadening of the decaying sterile neutrino. From Eq.~\eqref{eq:sterile_decay_width_degenerate}, we can see that $\Gamma_{N_1 \to N_2 \gamma} \propto |d_{NN\gamma}|^2 m_1^3$. After averaging over the observed field of view,
\begin{equation}
\frac{d\overline I_N}{d\lambda}
=
\frac{
f_N\,
\Gamma_{N_1\rightarrow N_2\gamma}
E_{\gamma,N}
}{
4\pi m_1
}
G_\lambda^{N}(\lambda)
\overline D_{\rm eff}(\lambda).
\label{eq:sterile_flux_averaged}
\end{equation}
The sterile-neutrino line intensity therefore scales as
\begin{equation}
I_N
\propto
f_N |d_{NN\gamma}|^2
\frac{(\Delta m)^4}{m_1},
\label{eq:sterile_flux_scaling}
\end{equation}
where from Eq.~\eqref{eq:sterile_decay_width_degenerate}, we used
$\Gamma_{N_1\rightarrow N_2\gamma}\propto |d_{NN\gamma}|^2(\Delta m)^3$
and $E_{\gamma,N}\simeq\Delta m$.

For the Majoron benchmark, each decay produces two photons with energy $E_{\gamma,\omega}=m_\omega/2$. The corresponding differential energy intensity is
\begin{equation}
I_{\omega}(\lambda_{\omega\gamma\gamma}) = \frac{dI_\omega}{d\lambda\,d\Omega}
=
\frac{
f_\omega\,
\Gamma_{\omega\rightarrow\gamma\gamma}
(2E_{\gamma,\omega})
}{
4\pi m_\omega
}
G_\lambda^\omega(\lambda)
D_{\rm eff}(\lambda,\ell,b),
\label{eq:majoron_flux}
\end{equation}
or, equivalently,
\begin{equation}
\frac{dI_\omega}{d\lambda\,d\Omega}
=
\frac{
f_\omega\,
\Gamma_{\omega\rightarrow\gamma\gamma}
}{
4\pi
}
G_\lambda^\omega(\lambda)
D_{\rm eff}(\lambda,\ell,b).
\label{eq:majoron_flux_simplified}
\end{equation}
Here $f_\omega$ is the Majoron dark matter fraction. The absence of the suppression factor $\Delta m/m_1$ distinguishes the Majoron two-photon signal from the sterile-neutrino transition signal. $G_\lambda^\omega(\lambda)$ represents the instrumental resolution and Doppler broadening of the decaying Majoron dark matter candidate. From Eq.~\eqref{eq:majoron_decay_width}, we can show that $\Gamma_{\omega\to\gamma\gamma}\propto \lambda_{\omega\gamma\gamma}^2 m_{\omega}^3$.

For later use, the two cases may be summarized by the unified expression
\begin{equation}
\frac{dI_X}{d\lambda\,d\Omega}
=
\frac{
f_X\Gamma_X N_{\gamma,X}E_{\gamma,X}
}{
4\pi m_X
}
G_\lambda^X(\lambda)
D_{\rm eff}(\lambda,\ell,b),
\label{eq:unified_flux}
\end{equation}
Here, $f_X$ is the abundance of dark matter carried by the decaying dark matter candidate $X$, $\Gamma_{X} $ is the decay width of the decaying dark matter candidate $X$, $E_{\gamma,X}$ is the photon energy of the decaying dark matter candidate $X$, $m_{X}$ is the mass of dark matter, $G_{\lambda}^X(\lambda)$ is the instrumental resolution and Doppler broadening of the decaying dark matter candidate $X$ and $D_{\rm eff}(\lambda, \ell, b)$ represents the effective dark matter density. where $X=N,\omega$. For the sterile transition,
\begin{equation}
N_{\gamma,N}=1,
\qquad
m_X=m_1,
\qquad
E_{\gamma,N}\simeq\Delta m,
\end{equation}
whereas for Majoron decay,
\begin{equation}
N_{\gamma,\omega}=2,
\qquad
m_X=m_\omega,
\qquad
E_{\gamma,\omega}=m_\omega/2.
\end{equation}
The subsequent JWST analysis uses this common line-flux formalism, with the sterile-neutrino transition magnetic moment treated as the
primary signal and the Majoron two-photon decay retained only as a benchmark comparison.

%\newpage
%=========================================================
\section{Results and Discussion}
\label{sec:results}
%=========================================================
\subsection{Sterile-to-sterile neutrino decay}
The James Webb Space Telescope (JWST) is a space-based infrared observatory and the successor to the \emph{Hubble Space Telescope}
~\cite{Gardner:2006ky,Rauscher:2007ta}. JWST provides continuous spectroscopic coverage over the wavelength range
\[
0.6~\mu{\rm m} \lesssim \lambda \lesssim 29~\mu{\rm m},
\]
making it particularly well suited for searches of narrow infrared spectral features from light, decaying dark matter candidates
~\cite{Gardner:2006ky}. In this work, we analyze publicly available JWST NIRSpec  Integral Field Unit (IFU) spectroscopic observations obtained from the MAST
archive.

The NIRSpec instrument employs dispersive optical elements to separate incident light into its constituent wavelengths
~\cite{saha2025shedding}. It is equipped with three diffraction gratings, \(\rm G140\), \(\rm G235\), and \(\rm G395\), as well as a
prism~\cite{jwst_nirspec_dispersers_2017,saha2025shedding}. Each grating supports both medium-resolution, \(R\sim1000\), and
high-resolution, \(R\sim2700\), spectroscopy~\cite{saha2025shedding}.
After dispersion, the light is recorded by the detector array, enabling spatially resolved spectroscopy over the Integral Field Unit (IFU) field of view.

For the present analysis, we focus exclusively on the \(\rm F170LP\)-\(\rm G235M\) filter-grating configuration. This choice is motivated by its wavelength coverage \cite{saha2025shedding},
\[
1.7~\mu{\rm m} \lesssim \lambda \lesssim 3.2~\mu{\rm m},
\]
which corresponds to infrared photon energies produced by the two-photon decay of dark
matter particles in the mass range $0.625~\rm eV\lesssim m_{\omega}\lesssim 1.17~\rm eV$, and this is relevant for the one-photon-decaying sterile-neutrino transition. The selected dataset has a
total integration time of \(1.897~{\rm ks}\), providing sufficient sensitivity to search for weak, narrow emission features.

Our analysis centers on IFU observations of the galaxy \(\rm GN\text{-}z11\), located at Galactic longitude \(\ell=126^\circ\) and latitude \(b=54.8^\circ\). Owing to its high Galactic latitude, foreground dust extinction along this line of sight is negligible~\cite{Schlegel:1997yv}, making it particularly well suited for indirect searches for dark matter decay signals in
the Milky Way halo.

Within this observational framework, we consider a keV-scale sterile neutrino as a decaying dark matter candidate residing in the Galactic
halo. We search for the characteristic monochromatic photon line arising from the radiative decay
\[
N_1\rightarrow N_2+\gamma,
\]
where \(m_1>m_2\) and \(\Delta m=m_1-m_2\). Using these data, we derive constraints on the magnetic coupling strength of the sterile-to-sterile transition magnetic moment, \(d_{NN\gamma}\).

We quantify the presence of a potential sterile-neutrino decay signal by constructing a \(\chi^2\) test statistic following
Refs.~\cite{saha2025shedding,Cowan:2010js,Roach:2022lgo},
\begin{equation}
\label{eq:chi2_sterile_result}
\chi^{2}(d_{NN\gamma}) =
\sum_i
\frac{
\left[
I_i-I_N(d_{NN\gamma})-S_{\alpha_i}(\lambda)
\right]^2
}
{\sigma_i^2},
\end{equation}
where \(I_i\) denotes the observed flux in the \(i\)th wavelength bin, $I_{N}(d_{NN\gamma})$ represents the differential flux of decaying sterile neutrino and \(\sigma_i\) represents the corresponding uncertainty. We adopt conservative uncertainties that are approximately \(2\)--\(3\) times larger than the nominal JWST instrumental errors, following
Ref.~\cite{janish2025hunting}. 

The observed fluxes \(I_i\) and uncertainties \(\sigma_i\) are taken from JWST IFU data obtained with the \(\rm F170LP\)-\(\rm G235M\) filter-grating configuration
~\cite{jwst_nirspec_dispersers_2017}. The smooth astrophysical continuum is modelled as a third-order polynomial, $S_{\alpha_i}(\lambda) = a + b\lambda + c\lambda^2 + d\lambda^3$, where $a$, $b$, $c$ and $d$ represent the coefficients of the polynomial ~\cite{mckinley1998cubic}. Here, by using the least-squares fitting procedure, we fit the \(a\), \(b\), \(c\), and
\(d\) coefficients, as well as the sterile-to-sterile transition magnetic coupling strength \(d_{NN\gamma}\), simultaneously such that the smooth astrophysical continuum is accounted for, and then we use the allowed $I_N(d_{NN\gamma})$ to derive the limit of sterile-to-sterile transition magnetic coupling strength $(d_{NN\gamma})$.  %The test statistic is minimized with respect to the spline coefficients \(a\), \(b\), \(c\), and
%\(d\), as well as the sterile-to-sterile transition magnetic coupling strength \(d_{NN\gamma}\), using a least-squares fitting procedure
%~\cite{mckinley1998cubic}. The best-fit coefficients are obtained by solving the corresponding Hessian matrix, yielding the minimum value \(\chi^2_{\rm min}\) for each trial mass splitting.

The statistical interpretation is performed in terms of
\[
\Delta\chi^2=\chi^2-\chi^2_{\rm min},
\]
with one degree of freedom~\cite{saha2025shedding}. Representative values of \(\Delta\chi^2\) corresponding to the \(1\sigma\) and
\(2\sigma\) confidence projections are extracted accordingly. We relegate the \(\chi^2\) scan as a function of \(\Delta m\) to
Appendix~\ref{app:deltachi2_mass} as a diagnostic of the fitting procedure~\cite{cochran1952chi2}.

To derive bounds on the sterile-to-sterile transition magnetic dipole coupling strength \(d_{NN\gamma}\), we employ the \(\chi^2\) test
statistic defined in Eq.~\eqref{eq:chi2_sterile_result}
~\cite{Cowan:2010js}. For each trial mass splitting, the \(\chi^2\) function is minimized with respect to the cubic spline coefficients
\(a\), \(b\), \(c\), and \(d\), as well as the transition magnetic coupling strength \(d_{NN\gamma}\), by solving the corresponding matrix equations.

We scan over \(1425\) trial values of the mass splitting between the two sterile-neutrino states in the range
\[
0.1~{\rm eV}\lesssim\Delta m\lesssim1~{\rm eV}.
\]
For each mass point, we extract the \(1\sigma\) and \(2\sigma\) confidence projections limits on \(d_{NN\gamma}\) by imposing the conditions
\(\Delta\chi^2\lesssim 1\) and \(\Delta\chi^2\lesssim 4\), respectively. The resulting confidence regions are obtained using the same test statistic as in Eq.~\eqref{eq:chi2_sterile_result}, where the
differential flux intensity \(I_N(d_{NN\gamma})\) is given in Eq.~\eqref{eq:sterile_flux}, \(S_{\alpha_i}(\lambda)\) denotes the smooth astrophysical continuum modelled by a cubic spline, and
\(\sigma_i\) represents the uncertainty of the JWST NIRSpec Integral Field Unit (IFU) spectroscopic data obtained with the \(\rm F170LP\)-\(\rm G235M\) filter-grating configuration~\cite{marston2018overview}. 

\begin{figure}[htbp]
    \centering
    \includegraphics[width=\linewidth]{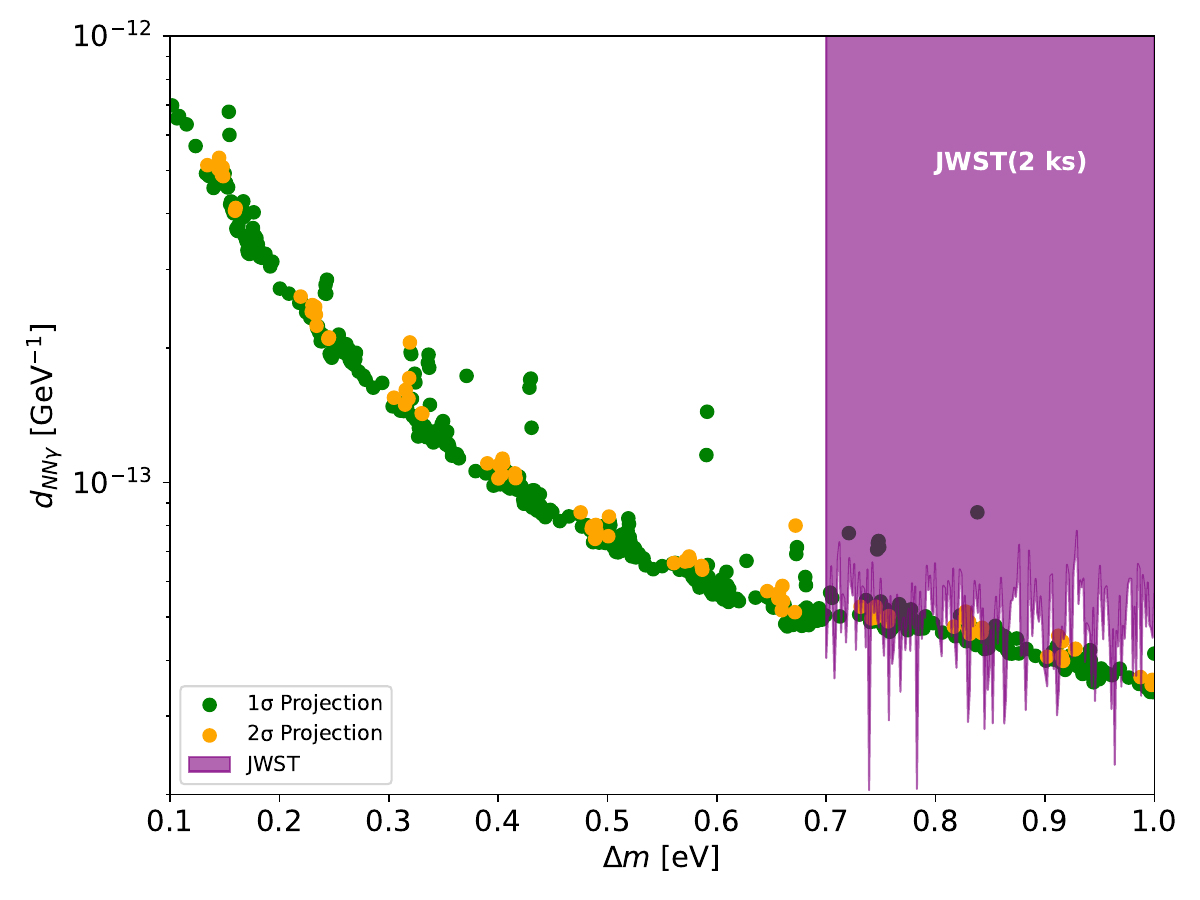}
    \caption{We present \(1\sigma\)--\(2\sigma\) projection upper limits on the sterile-to-sterile transition magnetic coupling
    strength \(d_{NN\gamma}\) in \({\rm GeV}^{-1}\) from an end-of-mission analysis of JWST NIRSpec Integral Field Unit (IFU) spectroscopic observations. We use the specific JWST filter \(\rm F170LP\) and grating
    \(\rm G235M\) for the mass-splitting range \(\Delta m\) between two sterile-neutrino states. The scattered green and yellow
    regions enclose the \(1\sigma\) and \(2\sigma\) projection limits, respectively. The purple region shows the JWST sensitivity.}
    \label{fig:sterile_coupling}
\end{figure}

The resulting \(1\sigma\) and \(2\sigma\) projection limits are shown in Fig.~\ref{fig:sterile_coupling} as scattered green and yellow regions, corresponding to \(\Delta\chi^2\lesssim 1\) and
\(\Delta\chi^2\lesssim 4\), respectively. For the JWST IFU dataset analysed here, we obtain constraints on the sterile-to-sterile transition magnetic coupling strength, $d_{NN\gamma}$ in the range
\[
10^{-14}~{\rm GeV}^{-1}
\lesssim
d_{NN\gamma}
\lesssim
10^{-11}~{\rm GeV}^{-1},
\]
for mass splittings in the range
\[
0.6~{\rm eV}
\lesssim
\Delta m
\lesssim
1~{\rm eV}.
\]
JWST has sensitivity to the mass-splitting range
\(0.1~{\rm eV}\lesssim\Delta m\lesssim1~{\rm eV}\). In Fig.~\ref{fig:sterile_coupling}, the projected \(1\sigma\)--\(2\sigma\) scattered region for \(d_{NN\gamma}\) is shown. The parameter space
below the purple JWST sensitivity curve is allowed, whereas above the purple region the predicted photon signal would lie above the JWST detection threshold.

\begin{figure}[htbp]
    \centering
    \includegraphics[width=\linewidth]{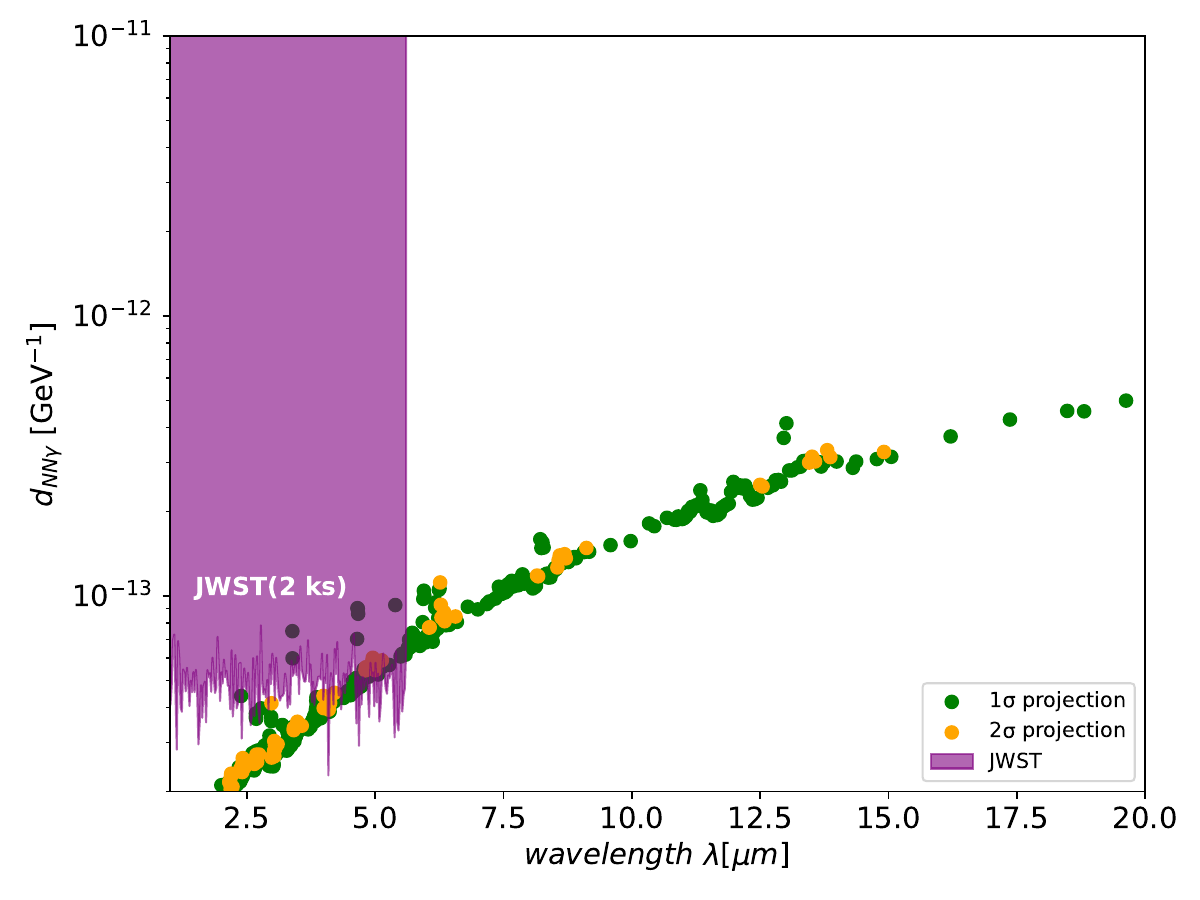}
    \caption{This figure illustrates the \(1\sigma\) and \(2\sigma\)  projection upper limits on the sterile-to-sterile transition
    magnetic dipole coupling strength \(d_{NN\gamma}\) in \({\rm GeV}^{-1}\) with respect to the photon wavelength
    \(\lambda\) in \(\mu{\rm m}\). The scattered green region  represents the \(1\sigma\) projection, whereas the scattered
    yellow region represents the \(2\sigma\) projection. The purple region shows the JWST sensitivity.}
    \label{fig:sterile_lambda}
\end{figure}

\begin{figure}[htbp]
    \centering
    \includegraphics[width=\linewidth]{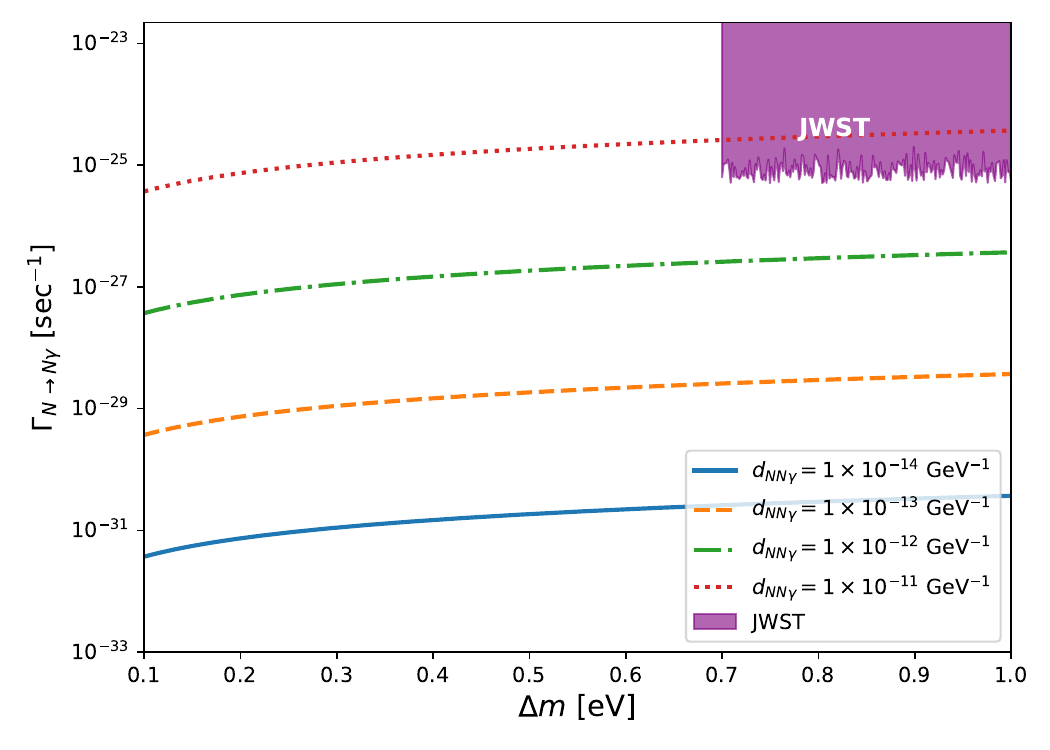}
    \caption{We illustrate the decay widths
    \(\Gamma_{N_1\to N_2\gamma}\) of one-photon decaying sterile  neutrinos versus the mass splitting \(\Delta m\) between two
    sterile-neutrino states in eV for the four representative values of the sterile-to-sterile transition magnetic coupling strength:
    \(d_{NN\gamma}=10^{-14}~{\rm GeV}^{-1}\),
    \(10^{-13}~{\rm GeV}^{-1}\), \(10^{-12}~{\rm GeV}^{-1}\), and \(10^{-11}~{\rm GeV}^{-1}\), respectively. The purple region denotes the JWST sensitivity region.}
    \label{fig:sterile_decay_width}
\end{figure}

\begin{figure}[htbp]
    \centering
    \includegraphics[width=0.5\textwidth]{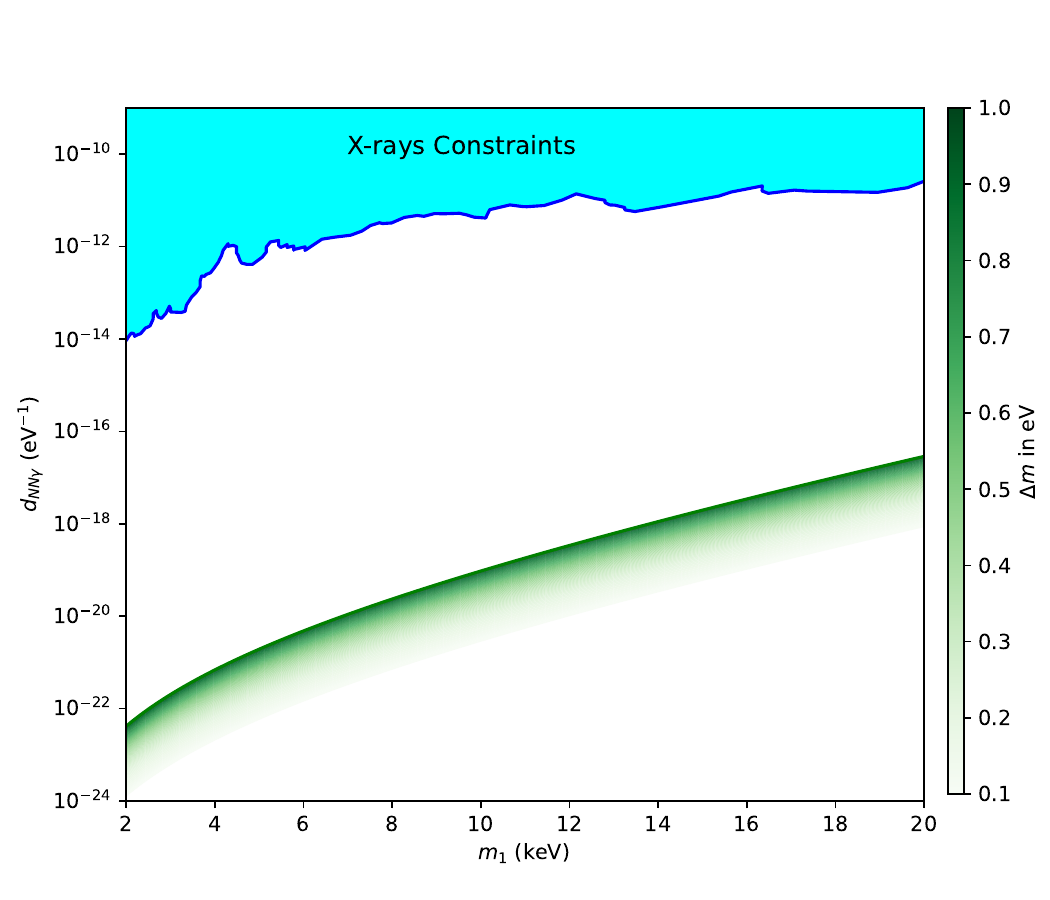}
    \caption{This figure illustrates the changes in the sterile-to-sterile transition magnetic coupling strength \(d_{NN\gamma}\) in \({\rm eV}^{-1}\) versus the mass of the
    sterile neutrino. The green band represents the changes of \(d_{NN\gamma}\) with respect to the sterile-neutrino mass obtained from the JWST flux, where \(\Delta m\) varies from
    \(0.1~{\rm eV}\) to \(1~{\rm eV}\). The cyan-shaded region represents the excluded region of \(d_{NN\gamma}-m_1\) values from X-ray observations.}
    \label{fig:dnn_m1}
\end{figure}

For the radiative decay channel \(N_1\to N_2\gamma\), each photon carries a monochromatic energy~\cite{saha2025shedding}
\[
E_\gamma=m_1-m_2=\Delta m.
\]
This corresponds to a photon wavelength~\cite{roy2025sensitivity}
\[
\lambda\simeq
1.24
\left(
\frac{1~{\rm eV}}{\Delta m}
\right)\mu{\rm m},
\]
as the mass splitting varies over the range
\(0.1~{\rm eV}\lesssim\Delta m\lesssim1~{\rm eV}\). The JWST wavelength coverage,
\[
0.6~\mu{\rm m}\lesssim\lambda\lesssim29~\mu{\rm m},
\]
corresponds to photon energies between approximately \(0.05~{\rm eV}\)
and \(2.1~{\rm eV}\), fully encompassing the infrared signatures expected from keV-mass-scale sterile-neutrino decays.
In Fig.~\ref{fig:sterile_lambda}, we present the resulting \(1\sigma\) and \(2\sigma\) upper limits on the sterile-to-sterile transition magnetic coupling strength \(d_{NN\gamma}\) as a function of the photon wavelength. These limits are derived from an end-of-mission sensitivity projection based on JWST NIRSpec Integral Field Unit (IFU) spectroscopic observations
and provide a complementary representation of the constraints obtained in the \((d_{NN\gamma},\Delta m)\) plane.
We further translate the constraints on the sterile-to-sterile transition magnetic coupling strength $(d_{NN\gamma})$ into bounds on the one-photon
decay width of the keV-mass-scale sterile-neutrino decay,\(\Gamma_{N_1\to N_2\gamma}\). The decay width is computed using Eq.~\eqref{eq:sterile_decay_width_delta}, evaluated for the four representative sterile-to-sterile transition magnetic coupling strength values

\[
d_{NN\gamma}
=
10^{-14},~10^{-13},~10^{-12},~10^{-11}
~{\rm GeV}^{-1},
\]
respectively, as shown in Fig.~\ref{fig:sterile_decay_width}.
As evident from Eq.~\eqref{eq:sterile_decay_width_delta}, the
radiative decay width scales with the cube of the mass splitting,
\[
\Gamma_{N_1\to N_2\gamma}\propto\Delta m^3.
\]
Consequently, for a fixed value of the coupling strength, the decay width increases monotonically across the mass-splitting range
\(0.1~{\rm eV}\lesssim\Delta m\lesssim1~{\rm eV}\). As shown in Fig.~\ref{fig:sterile_decay_width}, the cubic dependence of \(\Gamma_{N_1\to N_2\gamma}\) leads to a pronounced enhancement of
the decay width at larger \(\Delta m\), which directly underlies the strong JWST constraints in this region. This mass dependence reflects
the phase-space enhancement characteristic of two-body decays of sterile neutrinos and is an essential ingredient in interpreting the
JWST sensitivity to \(\Delta m\) in eV.
In Fig.~\ref{fig:sterile_decay_width}, we observe that the red dotted decay-width curve and the green dash-dotted decay-width curve corresponding to the sterile-to-sterile transition magnetic coupling
strengths \(10^{-11}~{\rm GeV}^{-1}\) and
\(10^{-12}~{\rm GeV}^{-1}\), respectively, lie within the JWST sensitivity region shown in purple, whose values extend up to \(10^{-25}~{\rm sec}^{-1}\) and \(10^{-27}~{\rm sec}^{-1}\),
respectively. In contrast, the orange dashed decay-width curve and the solid sky-blue decay-width curve corresponding to the coupling
strengths \(10^{-13}~{\rm GeV}^{-1}\) and
\(10^{-14}~{\rm GeV}^{-1}\) lie below the JWST sensitivity region, whose values reach up to \(10^{-29}~{\rm sec}^{-1}\) and
\(10^{-31}~{\rm sec}^{-1}\), respectively. We find that two of the decay-width curves fall within the JWST sensitivity region, and the
remaining two decay-width curves lie below the JWST sensitivity region. Our analysis allows us to investigate a substantial region of
the permissible parameter space for generic one-photon-decaying dark matter.
In Fig.~\ref{fig:dnn_m1}, we represent the sterile-to-sterile transition magnetic coupling strength \(d_{NN\gamma}\) in \({\rm GeV}^{-1}\) versus the mass of the decaying sterile neutrino \(m_1\) in the range
\[
1~{\rm keV}\lesssim m_1\lesssim20~{\rm keV},
\]
for the minimum required decay width of the sterile neutrino. The green colour band represents the changes in the sterile-to-sterile
transition magnetic moment \(d_{NN\gamma}\) for the JWST flux intensity with respect to the changes of the mass difference between two sterile neutrino states $\Delta m$ in $\rm eV$. From the decay-width equation
\eqref{eq:sterile_decay_width_delta}, we can see that the sterile-to-sterile transition magnetic moment directly depends on the decay width of the decaying sterile neutrino. We observe that, for the
small mass splitting \(\Delta m=0.1~{\rm eV}\), the
sterile-to-sterile transition magnetic coupling strength goes up to \(10^{-20}~{\rm GeV}^{-1}\). Whereas, for the larger mass splitting
\(\Delta m=1~{\rm eV}\), the sterile-to-sterile transition magnetic moment goes up to \(10^{-14}~{\rm GeV}^{-1}\). We show X-ray constraints on the sterile-to-sterile transition magnetic moment
\(d_{NN\gamma}\) in the cyan-shaded region.

The cyan-shaded region in Fig.~\ref{fig:dnn_m1} denotes existing
constraints on sterile-neutrino radiative decay from X-ray
observations. These bounds are most directly sensitive to photon
energies in the X-ray range, whereas the present JWST analysis probes
infrared photons generated by much smaller mass splittings,
\(\Delta m\sim0.1\)-\(1~{\rm eV}\). Therefore, the JWST limits are
not a direct replacement for X-ray searches, but rather provide a
complementary probe of the same sterile-to-sterile transition magnetic
interaction in a different photon-energy regime.

It is also useful to compare our result with constraints from the
cosmic optical background on radiatively decaying sterile neutrinos.
Such analyses typically probe larger transition energies and different
astrophysical datasets, while our study focuses on the JWST infrared
window. Within the parameter range shown in Fig.~\ref{fig:dnn_m1},
the JWST sensitivity provides an independent constraint on
\(d_{NN\gamma}\) for sterile-neutrino masses in the range
\(1~{\rm keV}\lesssim m_1\lesssim20~{\rm keV}\) and mass splittings
\(0.1~{\rm eV}\lesssim\Delta m\lesssim1~{\rm eV}\). This demonstrates
that infrared spectroscopy can probe a region of transition-magnetic-
moment parameter space that is complementary to both X-ray and optical
background searches.

\subsection{Sterile to active neutrino decay}
The
conventional sterile-to-active radiative decay \(N\to\nu\gamma\) is
controlled by active-sterile mixing, or by the active-to-sterile
transition dipole \(d_{N\nu\gamma}\), and the emitted photon energy is
set by the sterile-neutrino mass itself. For keV-scale sterile dark
matter this channel gives an X-ray line, while eV-scale sterile
neutrinos produced through active-sterile mixing, such as in the
Dodelson--Widrow mechanism, behave as hot or dark-radiation
components and are tightly constrained by CMB and large-scale
structure data~\cite{Planck:2018vyg,dodelson1994sterile}. More generally,
light thermal relics in the eV mass range are subject to stringent
cosmological limits from CMB anisotropies, weak lensing and galaxy
surveys~\cite{Xu:2021rwg},\cite{abbott2022dark}. These constraints apply to scenarios where
the sterile particle itself has an eV-scale mass. They do not directly
exclude the parameter space considered here, where the parent sterile
state is keV-scale and only the transition energy
\(\Delta m=m_1-m_2\) lies in the eV range. 
%=========================================================
\subsection{Minimal Majoron benchmark}
\label{subsec:majoron_result_minimal}
%=========================================================
%
For comparison, we also revisit the Majoron two-photon decay channel as a benchmark. In this case, an eV-scale Majoron dark matter candidate decays through
\(
\omega\rightarrow\gamma\gamma,
\)
producing a monochromatic infrared photon line in the JWST/NIRSpec wavelength range. Applying the same \(\chi^2\)-based JWST NIRSpec IFU spectroscopic data analysis to this channel, we obtain projected constraints on the
Majoron-photon coupling \(\lambda_{\omega\gamma\gamma}\).

\begin{figure}[htbp]
    \centering
    \includegraphics[width=\linewidth]{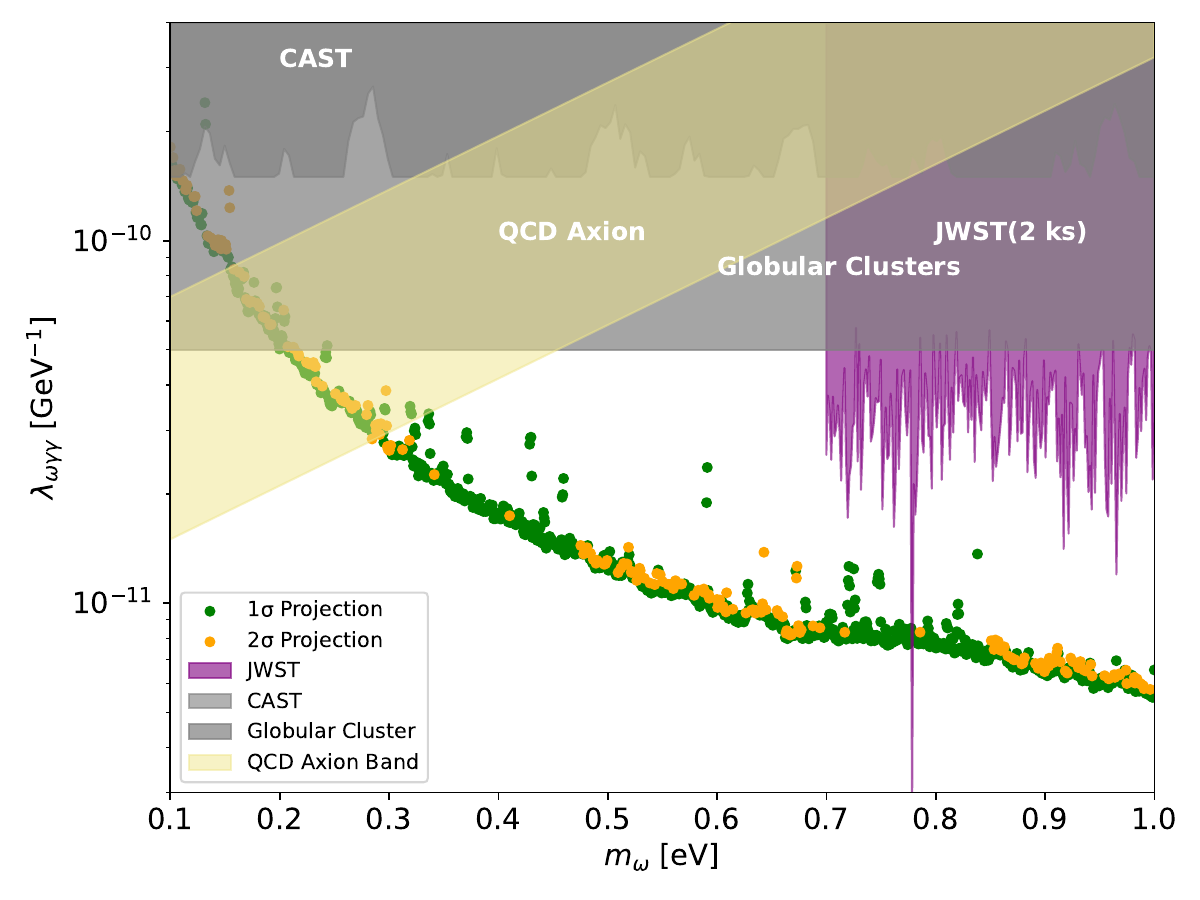}
    \caption{Projected \(1\sigma\) and \(2\sigma\) upper limits on
    the Majoron-photon coupling strength
    \(\lambda_{\omega\gamma\gamma}\) in \({\rm GeV}^{-1}\) as a function of the Majoron mass \(m_\omega\). The green and orange
    points represent the \(1\sigma\) and \(2\sigma\) projections, respectively, obtained using the JWST/NIRSpec IFU spectroscopic
    \(\rm F170LP\) configuration. The purple region denotes the JWST sensitivity, while the grey shaded regions show existing constraints from CAST and globular clusters. The yellow band indicates the QCD-axion parameter region.}
    \label{fig:majoron_coupling}
\end{figure}

For the JWST NIRSpec IFU spectroscopic dataset analysed here, the resulting \(1\sigma\) and
\(2\sigma\) projection limits are shown in
Fig.~\ref{fig:majoron_coupling}. We obtain constraints on the Majoron-photon coupling strength $(\lambda_{\omega\gamma\gamma})$ in the range
\[
10^{-11}~{\rm GeV}^{-1}
\lesssim
\lambda_{\omega\gamma\gamma}
\lesssim
10^{-9}~{\rm GeV}^{-1},
\]
for Majoron masses in the range
\[
0.1~{\rm eV}
\lesssim
m_\omega
\lesssim
1~{\rm eV}.
\]
This result is included as a benchmark comparison, since the Majoron two-photon signal has the same line phenomenology as the standard axion-like-particle two-photon decay interpretation.

For the radiative decay channel \(\omega\to\gamma\gamma\), each
photon carries a monochromatic energy~\cite{saha2025shedding}
\(
E_\gamma=\frac{m_\omega}{2}.
\)
This corresponds to a photon wavelength~\cite{roy2025sensitivity}
\[
\lambda\simeq
2.48
\left(
\frac{1~{\rm eV}}{m_\omega}
\right)\mu{\rm m},
\]
as the Majoron mass varies over the range
\(0.1~{\rm eV}\lesssim m_\omega\lesssim1~{\rm eV}\). This provides a direct comparison with the sterile-neutrino transition case, where
the photon energy is controlled by the mass splitting \(\Delta m\).

\begin{figure}[htbp]
    \centering
    \includegraphics[width=\linewidth]{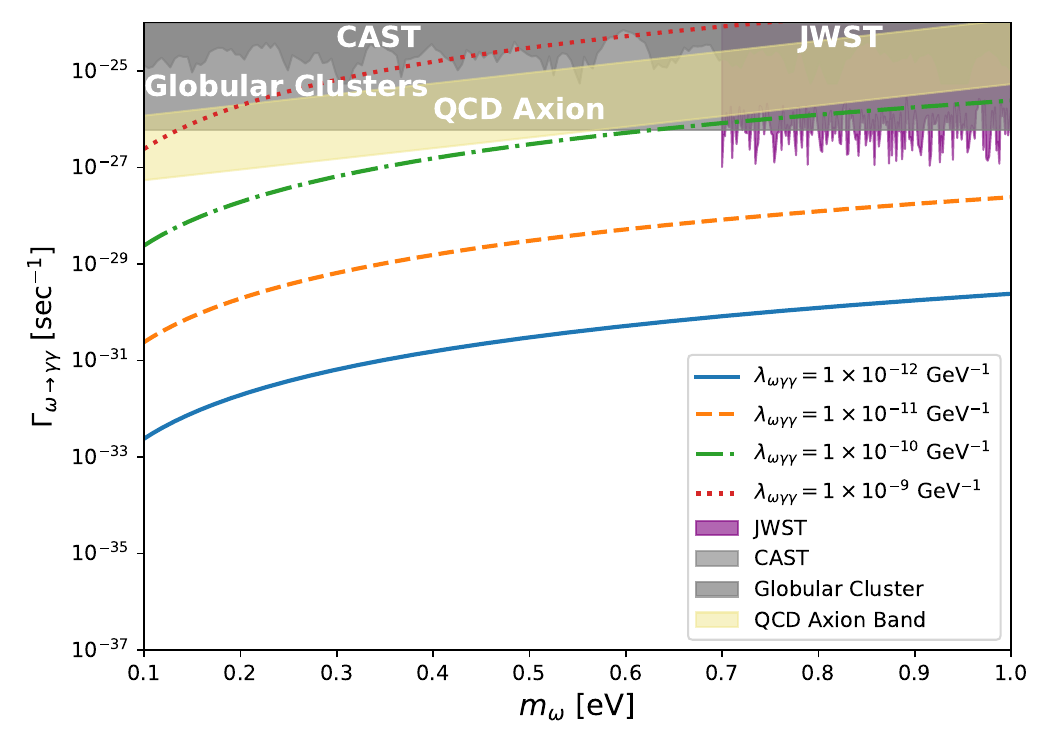}
    \caption{We illustrate the decay width
    \(\Gamma_{\omega\to\gamma\gamma}\) of two-photon decaying Majorons versus the Majoron mass \(m_\omega\) in eV for the four
    representative values of the Majoron-photon coupling strength:\(\lambda_{\omega\gamma\gamma}=10^{-12}~{\rm GeV}^{-1}\),\(10^{-11}~{\rm GeV}^{-1}\), \(10^{-10}~{\rm GeV}^{-1}\), and
    \(10^{-9}~{\rm GeV}^{-1}\), respectively. The purple region denotes the JWST sensitivity region. We depict the existing constraints on the decay width of decaying Majorons by CAST, globular clusters, and QCD axions.}
    \label{fig:majoron_decay_width}
\end{figure}

We further translate the constraints on the Majoron-photon coupling $\lambda_{\omega\gamma\gamma}$ into bounds on the two-photon decay width \(\Gamma_{\omega\to\gamma\gamma}\). The decay width is computed using Eq.~\eqref{eq:majoron_decay_width}, evaluated for the four representative Majoron-photon coupling strengths
\[
\lambda_{\omega\gamma\gamma}
=
10^{-12},~10^{-11},~10^{-10},~10^{-9}
~{\rm GeV}^{-1}.
\]
For a fixed value of \(\lambda_{\omega\gamma\gamma}\), the decay width
increases monotonically with \(m_\omega\), reflecting the cubic mass dependence of the two-body decay.
The corresponding Majoron-photon coupling sensitivity is shown in
Fig.~\ref{fig:majoron_coupling}. The green and orange points denote
the projected \(1\sigma\) and \(2\sigma\) upper limits obtained from
the JWST IFU analysis, respectively. The projected sensitivity
improves toward larger Majoron masses, reaching the level of
\(\lambda_{\omega\gamma\gamma}\sim{\rm few}\times10^{-12}\,
{\rm GeV}^{-1}\) near \(m_\omega\simeq1~{\rm eV}\). This behaviour
follows from the cubic dependence of the two-photon decay width on
the Majoron mass, which enhances the expected line intensity at
larger \(m_\omega\) for a fixed coupling strength.
For comparison, we also show existing constraints from CAST, globular
clusters, and the QCD-axion band. The JWST sensitivity shown in
purple becomes relevant in the higher-mass part of the range
considered here, approximately
\(0.7~{\rm eV}\lesssim m_\omega\lesssim 1~{\rm eV}\). In this region,
the projected JWST reach probes Majoron-photon couplings below the
CAST and globular-cluster-excluded regions, demonstrating that
infrared line searches provide a complementary probe of eV-scale
pseudoscalar dark matter.
As shown in Fig.~\ref{fig:majoron_decay_width}, the red dotted and green dash-dotted curves corresponding to
\(\lambda_{\omega\gamma\gamma}=10^{-9}~{\rm GeV}^{-1}\) and \(10^{-10}~{\rm GeV}^{-1}\), respectively, lie within the JWST sensitivity region shown in purple. Their decay widths reach values up to approximately \(10^{-24}~{\rm sec}^{-1}\) and
\(10^{-26}~{\rm sec}^{-1}\), respectively. In contrast, the orange dashed and solid sky-blue curves corresponding to \(\lambda_{\omega\gamma\gamma}=10^{-11}~{\rm GeV}^{-1}\) and \(10^{-12}~{\rm GeV}^{-1}\) lie below the JWST sensitivity region,
with decay widths reaching approximately \(10^{-28}~{\rm sec}^{-1}\)
and \(10^{-30}~{\rm sec}^{-1}\), respectively. We find that two of the Majoron decay-width curves fall within the JWST sensitivity region, and the remaining two curves lie below it. In Fig.~\ref{fig:majoron_coupling}, we have shown the existing constraints on the Majoron-photon coupling strength $\lambda_{\omega\gamma\gamma}~\rm in ~GeV^{-1}$ with respect to the Majoron mass $m_{\omega}$ in eV. As from Eq.~\eqref{eq:majoron_decay_width}, we observe that the decay width of decaying majoron-like particles is proportional to the square of the majoron-photon coupling strength $\Gamma_{\omega\to \gamma\gamma}\propto\lambda_{\omega\gamma\gamma}^{2}$. So, by substituting the $\lambda_{\omega\to\gamma\gamma}$ values in Eq.~\eqref{eq:majoron_decay_width}, we illustrate the existing constraints on the decay width of decaying majorons into two photons $\Gamma_{\omega\to\gamma\gamma}$ in Fig.~\ref{fig:majoron_decay_width}. Here, we find that the decay width of decaying Majorons for the Majoron-photon coupling strength $10^{-9}~\rm GeV^{-1}$ falls within the existing constraint regions, so we can exclude this. Whereas we can include the decay width of decaying Majorons for three Majoron-photon coupling strengths, $\lambda_{\omega\gamma\gamma} = [10^{-10}, 10^{-11}, 10^{-12}]~\rm GeV^{-1},$respectively. This minimal
Majoron result illustrates the reach of JWST for generic two-photon-decaying dark matter, while the main focus of this work remains the
sterile-to-sterile transition magnetic moment.

\subsection{Common neutrino-sector interpretation}
\label{subsec:common_neutrino_sector}
%=========================================================

Although the sterile-neutrino transition magnetic moment and the Majoron two-photon decay arise from different effective interactions, they represent two closely related infrared-line signatures of
physics beyond the Standard Model neutrino sector. Sterile neutrinos are naturally motivated by neutrino-mass generation, while Majorons
arise in theories where global lepton number is spontaneously broken. Therefore, a broad class of BSM neutrino-sector frameworks can contain
both quasi-degenerate sterile states and light pseudoscalar degrees of freedom.
The important phenomenological point is that both possibilities yield narrow photon lines within the JWST wavelength range. In the sterile-neutrino case, the relevant signal is the one-photon
transition
\begin{equation}
N_1\rightarrow N_2+\gamma,
\end{equation}
induced by the sterile-to-sterile transition magnetic coupling $d_{NN\gamma}$. The photon energy is controlled by the small mass splitting \(\Delta m=m_1-m_2\), so that keV-scale sterile-neutrino
dark matter can still produce an eV-scale infrared photon. In the Majoron case, the corresponding signal is the two-photon decay
\(
\omega\rightarrow\gamma\gamma,
\)
controlled by the effective coupling \(\lambda_{\omega\gamma\gamma}\),
with the photon energy fixed by the Majoron mass.
Thus, JWST does not only test a specific dark matter candidate; it probes a class of neutrino-sector interactions capable of producing
infrared lines. The sterile-neutrino transition magnetic moment is the main new focus of this work, because its kinematics differ from the
standard axion-like or Majoron-to-two-photon-decay interpretation. The Majoron result is included as a minimal benchmark to show that the same JWST
line-search framework can also be applied to pseudoscalar states associated with lepton-number breaking.

In a complete ultraviolet theory, these two effects may be connected. For example, spontaneous lepton-number breaking can generate a Majoron and simultaneously motivate a pseudo-Dirac sterile-neutrino sector, while charged mediator fields can induce the transition magnetic moment. We do not require such a model-specific connection
in the present analysis. Instead, we treat \(d_{NN\gamma}\) and \(\lambda_{\omega\gamma\gamma}\) as independent effective parameters and use JWST data to illustrate the infrared reach for both classes
of neutrino-sector dark matter signals.
This comparison shows that JWST infrared spectroscopy is a useful probe of BSM neutrino physics beyond the usual X-ray sterile-neutrino
searches. It can constrain not only radiative transitions among sterile states, but also photon-line signals from light Majoron-like particles. The two cases should therefore be viewed as complementary manifestations of a broader neutrino-sector dark matter program, rather than as two unrelated studies.

%\newpage

%=========================================================
\section{Ultraviolet interpretation}
\label{sec:UV_interpretation}
%=========================================================

The limits derived in this work are presented in terms of the low-energy sterile-to-sterile transition magnetic coupling \(d_{NN\gamma}\). Nevertheless, it is useful to indicate how such an
operator may arise from a simple ultraviolet completion. This discussion is not required for the JWST recast itself, but it provides a concrete interpretation of the effective parameter constrained in the previous section.
Following the representative construction of
Ref.~\cite{Beltran:2024twr}, the transition dipole can be generated at one loop by introducing a vector-like charged lepton \(E\) and a
complex charged scalar \(\phi\), with gauge quantum numbers
\begin{equation*}
E_{L,R}\sim(\mathbf{1},\mathbf{1},-1),
\qquad
\phi\sim(\mathbf{1},\mathbf{1},-1),
\label{eq:charged_mediator_quantum_numbers}
\end{equation*}
under \(SU(3)_c\times SU(2)_L\times U(1)_Y\). The relevant mediator Lagrangian is
\begin{align}
\mathcal{L}_{\rm med}\supset{}&
\overline{E}\left(i\slashed{D}-m_E\right)E
+
(D_\mu\phi)^\dagger(D^\mu\phi)
-
m_\phi^2|\phi|^2 \nonumber \\
&-
\lambda_{\phi H}|\phi|^2(H^\dagger H)
\nonumber
-
\left[
h_i\,\overline{N_{Ri}}E_L\phi^\ast
\right . \\& \left .+
h'_i\,\overline{N_{Ri}^{\,c}}E_R\phi^\ast
+
{\rm H.c.}
\right],
\label{eq:charged_mediator_lagrangian}
\end{align}
where \(i=1,2\) labels the sterile-neutrino mass eigenstates, ordered as \(m_1>m_2\). The two chiral couplings \(h_i\) and \(h'_i\) are both
needed to generate the chirality-flipping electromagnetic transition
dipole.

After integrating out \(E\) and \(\phi\), the ultraviolet theory generates the dimension-five \(N_R\)SMEFT operator
\begin{equation}
\mathcal{O}_{NNB}^{(5)ij}
=
\left(
\overline{N_{Ri}^{\,c}}\sigma^{\mu\nu}N_{Rj}
\right)B_{\mu\nu}.
\label{eq:ONNB_operator}
\end{equation}
For \(m_{1,2}\ll m_E,m_\phi\), the corresponding one-loop matching coefficient is~\cite{Beltran:2024twr}
\begin{equation}
C_{NNB}^{(5)ij}
=
\frac{g'}{64\pi^2m_E}
\left(
h'_i h_j^\ast-h_i^\ast h'_j
\right)
\mathcal{F}(r),
\qquad
r\equiv\frac{m_\phi^2}{m_E^2},
\label{eq:CNNB_matching}
\end{equation}
where
\begin{equation}
\mathcal{F}(r)
=
\frac{1}{1-r}
+
\frac{r\ln r}{(1-r)^2},
\qquad
\mathcal{F}(1)=\frac{1}{2}.
\label{eq:dipole_loop_function}
\end{equation}
After electroweak symmetry breaking, \(B_{\mu\nu}=c_WF_{\mu\nu}-s_WZ_{\mu\nu}\). Therefore, the photon transition coefficient is \cite{Beltran:2024twr}
\begin{equation}
d_{NN\gamma}^{ij}
=
c_W C_{NNB}^{(5)ij}.
\label{eq:photon_matching}
\end{equation}
For the decay \(N_1\to N_2\gamma\), we denote the relevant physical coefficient by \cite{Beltran:2024twr}
\begin{equation}
d_{NN\gamma}
\equiv
d_{NN\gamma}^{21}
=
\frac{e}{64\pi^2m_E}
\left(
h'_2h_1^\ast-h_2^\ast h'_1
\right)
\mathcal{F}(r),
\label{eq:dNNgamma_matching}
\end{equation}
where \(e=g'c_W\). The overall sign of this coefficient is not physical for the decay rate, since the observable width depends on
\(|d_{NN\gamma}|^2\). The antisymmetric flavour structure in Eq.~\eqref{eq:dNNgamma_matching} also reflects the fact that sterile neutrinos can have transition dipole moments, while diagonal
electromagnetic dipole moments vanish.

At energies below the electroweak and mediator scales, the matching reduces to the \(N_R\)LEFT interaction used in the phenomenological
analysis,
\begin{equation}
\mathcal{L}_{N_R{\rm LEFT}}
\supset
d_{NN\gamma}
\left(
\overline{N_2^{\,c}}\sigma^{\mu\nu}P_RN_1
\right)F_{\mu\nu}
+
{\rm H.c.},
\label{eq:low_energy_dNNgamma}
\end{equation}
which induces the radiative decay \(N_1\to N_2\gamma\).
A complete ultraviolet model must also ensure that loop corrections to the sterile-neutrino mass matrix do not destabilize the eV-scale
splitting \(\Delta m=m_1-m_2\). This can be achieved through an appropriate sterile-flavour structure, alignment, or approximate
symmetry. In the present work we do not impose a specific UV flavour model. Instead, Eq.~\eqref{eq:dNNgamma_matching} is used only to show that the effective coupling \(d_{NN\gamma}\) constrained by JWST can arise from a simple charged-mediator completion.
For comparison, a light Majoron-like pseudoscalar can also be embedded in a broad neutrino-sector ultraviolet framework. In Majoron models,
a complex scalar field spontaneously breaks global lepton number, and the Majoron appears as the associated pseudo--Nambu--Goldstone boson
\cite{Chikashige:1980qk,Gelmini:1980re,Schechter:1981cv}. If additional charged states or anomaly-like interactions generate an effective coupling to photons, the low-energy theory contains
\begin{equation}
\mathcal L_{\omega}^{\rm eff}
\supset
\frac{1}{4}
\lambda_{\omega\gamma\gamma}
\omega F_{\mu\nu}\widetilde F^{\mu\nu},
\label{eq:majoron_uv_eff}
\end{equation}
which induces \(\omega\to\gamma\gamma\). Such anomalous Majoron realizations provide a useful benchmark for infrared line searches.
However, in this work the Majoron result is included only for comparison; the main UV interpretation and the primary JWST limits
concern the sterile-to-sterile transition magnetic moment
\(d_{NN\gamma}\).
%=========================================================
\section{Conclusion and Outlook}
\label{sec:Conclusion}
%=========================================================

In this work, we have investigated the prospects for probing
keV-scale decaying sterile-neutrino dark matter using infrared
spectroscopy with the James Webb Space Telescope. Our analysis is
based on publicly available blank-sky data from the JWST NIRSpec Integral Field Unit (IFU)
instrument, employing the \(\rm F170LP\) filter in combination with
the \(\rm G235M\) grating~\cite{marston2018overview}. These
observations provide a unique opportunity to search for narrow
infrared emission lines arising from the radiative decay of
quasi-degenerate sterile-neutrino dark matter states.
Focusing on JWST NIRSpec Integral Field Unit (IFU) observations toward the high-Galactic-latitude target
\(\rm GN\text{-}z11\), where foreground extinction is negligible, we
performed a model-independent spectral analysis to search for
monochromatic photon lines in the JWST data. By constructing a
\(\chi^2\)-based test statistic and modelling the astrophysical
continuum with a cubic spline, we derived sensitivity regions for the
sterile-to-sterile transition magnetic coupling strength
\(d_{NN\gamma}\) over the mass splitting between two sterile-neutrino
states. We then translated these limits into corresponding bounds on
the one-photon decay width \(\Gamma_{N_1\to N_2\gamma}\).

For mass splittings in the range,
\[
0.1~{\rm eV}\lesssim \Delta m \lesssim 1~{\rm eV},
\]
our analysis constrains the radiative decay
\(N_1\to N_2\gamma\), where the emitted photon carries energy
\(E_\gamma\simeq \Delta m\). We find that the decay-width curves
corresponding to the larger representative transition magnetic
couplings fall within the JWST sensitivity region, while the curves
corresponding to smaller couplings lie below the present sensitivity.
The decay-width behaviour remains consistent with the observational
limits established by the JWST NIRSpec Integral Field Unit (IFU) spectroscopic dataset.
Our strongest sensitivity reaches approximately
\[
d_{NN\gamma}
\lesssim
7\times10^{-14}~{\rm GeV}^{-1},
\]
the corresponding sterile-neutrino one-photon decay width,
reaching the level of
\[
\Gamma_{N_1\to N_2\gamma}
\lesssim
10^{-25}~{\rm sec}^{-1},
\]
the most sensitive part of the mass-splitting range considered in
this work. These results show that JWST infrared spectroscopy can
probe sterile-to-sterile transition magnetic moments in a region
complementary to conventional X-ray, laboratory and stellar searches.
For comparison, we have also revisited the Majoron two-photon decay
channel as a minimal benchmark. Majorons arise as pseudo-Nambu-Goldstone bosons associated with the spontaneous
breaking of global lepton number and are naturally connected to
neutrino-mass generation mechanisms~\cite{hill2007neutrino}. In the
eV mass range, a Majoron-like dark matter candidate can produce an
infrared line through the decay \(\omega\to\gamma\gamma\), with each
photon carrying energy \(E_\gamma=m_\omega/2\). Applying the same
JWST/NIRSpec IFU line-search strategy, we obtain projected
constraints on the Majoron-photon coupling
\(\lambda_{\omega\gamma\gamma}\) in the range
\[
10^{-11}~{\rm GeV}^{-1}
\lesssim
\lambda_{\omega\gamma\gamma}
\lesssim
10^{-9}~{\rm GeV}^{-1},
\]
for Majoron masses around
\[
0.6~{\rm eV}\lesssim m_\omega\lesssim1~{\rm eV}.
\]
The corresponding two-photon decay-width curves show that the larger
representative couplings fall within the JWST sensitivity region,
whereas the smaller couplings remain below the present sensitivity.
This Majoron result is included as a benchmark comparison, while the
main focus of the present work remains the sterile-to-sterile
transition magnetic moment \(d_{NN\gamma}\).
Overall, our results demonstrate that JWST infrared spectroscopy
provides a powerful and complementary probe of decaying dark matter
signals associated with the broader BSM neutrino sector. The sterile
transition magnetic moment and the Majoron two-photon decay represent
two different mechanisms that can produce narrow infrared photon
lines, and both can be tested using blank-sky JWST observations. With
increased exposure times, additional blank-sky fields, improved sky
coverage and a dedicated treatment of instrumental systematics,
future JWST datasets are expected to improve the sensitivity to
\(d_{NN\gamma}\), \(\Gamma_{N_1\to N_2\gamma}\), and related
neutrino-sector dark matter line signals. Furthermore, we can conclude that JWST observations provide a more stringent constraint on the sterile-to-sterile transition magnetic coupling strength $(d_{NN\gamma})$ than the anomalous cosmic optical background (COB) measurement due to sterile-to-sterile neutrino decay.
\section{Acknowledgement}
TS is supported by the DST, Government of India, through the DST
INSPIRE Faculty Fellowship (DST/INSPIRE/04/2024/004616).
We acknowledge the Mikulski Archive for Space Telescopes (MAST), from which all James Webb Space Telescope (JWST) data used in this work were obtained \cite{marston2018overview}. All data are available at: 
\href{https://mast.stsci.edu/portal/Mashup/Clients/Mast/Portal.html?searchQuery=%7B%22service%22:%22DOIOBS%22,%22inputText%22:%2210.17909/3e5f-nv69%22%7D}
{https://archive.stsci.edu/doi/}.

%%%%%%%%%%%%%%%%%%%%%%%%%%%%%%%%%%%%%%%%%%%%%%%

%%%%%%%%%%%%%%%%%%%%%%%%%%%%%%%%%%%%%%%%%%%%%%%

\bibliographystyle{apsrev4-2}
\bibliography{mainbib}
%\newpage

\appendix
\label{appendix}
\section{Goodness of fit}
\label{app:deltachi2_mass}

\begin{figure}[h]
    \centering
    \includegraphics[width=0.8\linewidth]{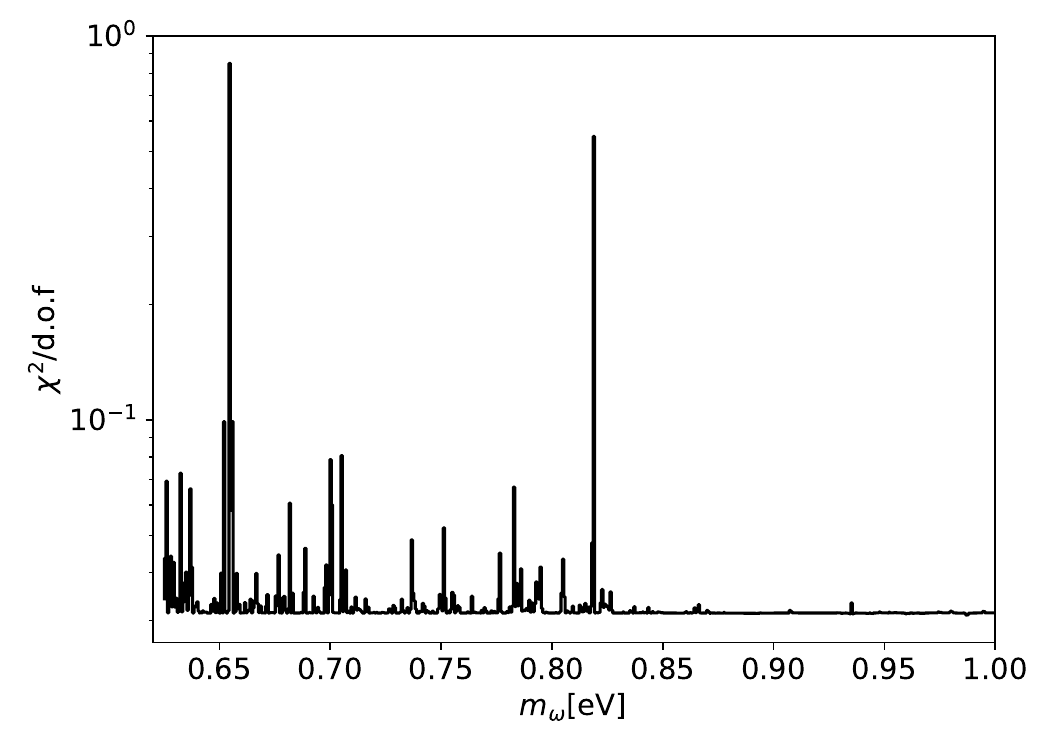}
    \caption{Logarithmic variation of the $\chi^2$ value per degree of freedom as a function of the Majoron mass $m_\omega$. }
    \label{fig:chi2_mass_appendix}
\end{figure}

Figure~\ref{fig:chi2_mass_appendix} shows the logarithm of $\chi^2$ per degree of freedom as a function of the Majoron mass $m_\omega$. The plot shows the robustness and consistency check of the fitting procedure. The physically relevant constraints on the decaying Majoron scenario are instead presented in the main text in the form of exclusion limits in the $(\lambda_{\omega\gamma\gamma},\, m_\omega)$ plane at fixed confidence levels.

\end{document}